\documentclass[twoside,twocolumn,10pt]{article}

\usepackage{wscg}           
\RequirePackage{ifpdf}
\ifpdf
 \RequirePackage[pdftex]{graphicx}
 \RequirePackage[pdftex]{color}
\else
 \RequirePackage[dvips,draft]{graphicx}
 \RequirePackage[dvips]{color}
\fi

\title{SEL-CIE: Knowledge-Guided Self-Supervised Learning Framework for CIE-XYZ Reconstruction from Non-Linear sRGB Images}

\author{
\makebox[\textwidth]{%
\parbox{0.22\textwidth}{\centering
Shir Barzel\\[1mm]
Tel Aviv University\\
sbarzel@gmail.com
}
\hfill
\parbox{0.22\textwidth}{\centering
Moshe Salhov\\[1mm]
Playtika LTD\\
moshesa@playtika.com
}
\hfill
\parbox{0.22\textwidth}{\centering
Ofir Lindenbaum\\[1mm]
Bar Ilan University\\
ofirlin@gmail.com
}
\hfill
\parbox{0.22\textwidth}{\centering
Amir Averbuch\\[1mm]
Tel Aviv University\\
amir1@tauex.tau.ac.il
}
}
}

\usepackage{url}
\makeatletter
\def\Uslash{\mathbin{\mathchar`\/}\@ifnextchar{/}{\kern-.15em}{}}
\g@addto@macro\UrlSpecials{\do \/ {\Uslash}}
\def\Ucolon{\mathbin{\mathchar`:}\@ifnextchar{/}{\kern-.1em}{}}
\g@addto@macro\UrlSpecials{\do : {\Ucolon}}
\makeatother

\begin{document}

\twocolumn[{\csname @twocolumnfalse\endcsname

\maketitle  

\begin{abstract}
\noindent

Modern cameras typically offer two types of image states: a minimally processed linear raw RGB image representing the raw sensor data, and a highly-processed non-linear image state, such as the sRGB state. The CIE-XYZ color space is a device-independent linear space used as part of the camera pipeline and can be helpful for computer vision tasks, such as image deblurring, dehazing, and color recognition tasks in medical applications, where color accuracy is important. However, images are usually saved in non-linear states, and achieving CIE-XYZ color images using conventional methods is not always possible. To tackle this issue, classical methodologies have been developed that focus on reversing the acquisition pipeline. More recently, supervised learning has been employed, using paired CIE-XYZ and sRGB representations of identical images. However, obtaining a large-scale dataset of CIE-XYZ and sRGB pairs can be challenging. To overcome this limitation and mitigate the reliance on large amounts of paired data, self-supervised learning (SSL) can be utilized as a substitute for relying solely on paired data. This paper proposes a framework for using SSL methods alongside paired data to reconstruct CIE-XYZ images and re-render sRGB images, outperforming existing approaches. The proposed framework is applied to the sRGB2XYZ dataset.
\end{abstract}

\subsection*{Keywords}
CIE-XYZ Color Space, sRGB, Image Reconstruction, Self-Supervised Learning (SSL), Raw Image, Macbeth ColorChecker
\vspace*{1.0\baselineskip}
}]


\section{Introduction}


In the realm of digital photography, a customary process involves the transformation of a sensor RAW image captured by a digital camera into the standardized sRGB format utilizing an in-camera Image Signal Processor (ISP) \cite{karaimer2016software}. Traditional ISPs are optimized primarily to generate visually appealing, compressed RGB images that cater to human perception. The pervasive availability of such RGB images on the internet has contributed significantly to the recent advancements in machine learning-based computer vision technologies.
The image processor in a digital camera applies various adjustments to the captured sensor image \cite{brown2019understanding}. In the initial stage, linear operations like white balance and color adaptation transform the sensor-specific raw RGB image into a standardized color space, such as CIE-XYZ \cite{kerr2010cie}. This creates a scene-referred image that directly correlates with the original captured scene. Subsequently, the "photo-finishing" stage involves applying non-linear adjustments and local operators to enhance the visual aesthetics of the photograph. This may include selectively manipulating colors to improve skin tones or increasing local contrast for a more striking appearance. Finally, the processed image is converted to the desired output color space.

 The increasing prevalence of digital imaging has propelled the development of modern cameras that provide users access to either one of the two distinct image states: minimally processed linear raw RGB data and highly processed non-linear images, such as those in the sRGB state.
 These two image states serve different purposes, with the former representing the raw sensor data and the latter addressing the visual perception of users.
With its linear relationship to scene radiance, the raw-RGB image state offers advantages for low-level computer vision tasks such as deblurring, dehazing, denoising, and image enhancement \cite{tai2013nonlinear,brooks2019unprocessing,zamir2020cycleisp}. However, the sensor-specific nature of color filter arrays in the raw RGB format leads to significant variations in captured values between different sensors, often necessitating sensor or camera-specific tailor-made algorithms.
The display-referred image state, typically in the sRGB color space, is widely used for display purposes but can vary significantly in value due to proprietary photo-finishing applied by different cameras. This leads to differences in sRGB values between images captured of the same scene using different camera models or settings.

\begin{figure*}[t]
\centering
\fbox{
\begin{tabular}{ccc}
   \includegraphics[width=0.25\linewidth]{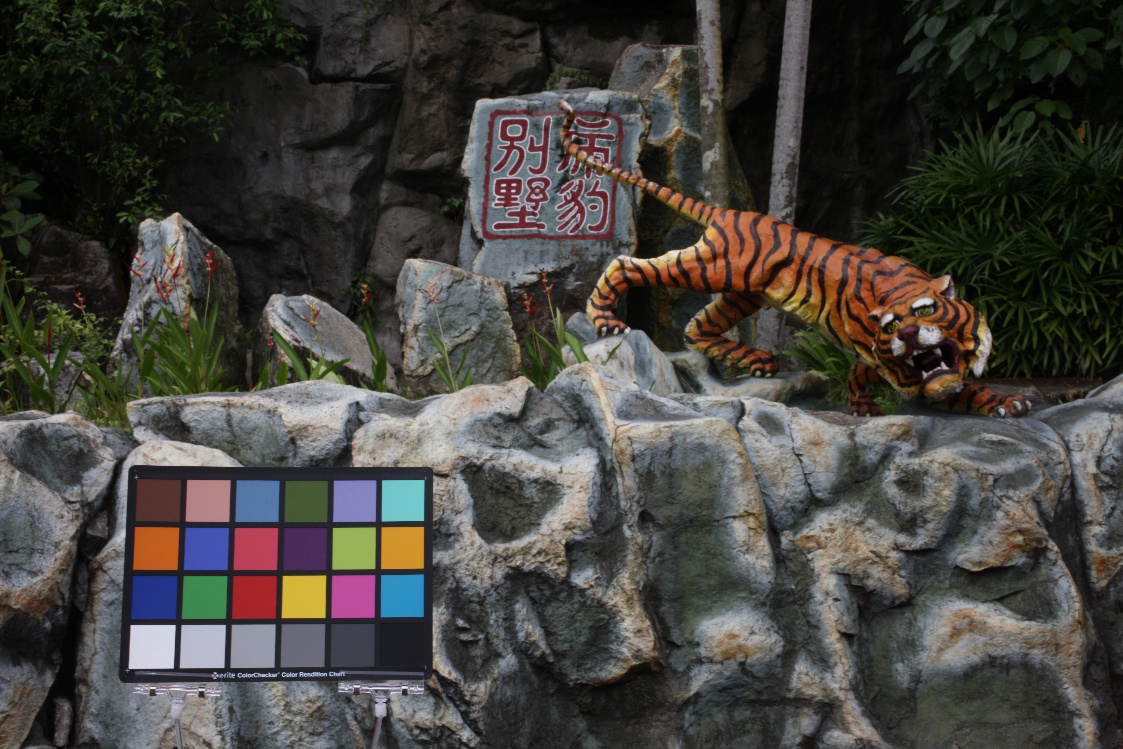} & \includegraphics[width=0.25\linewidth]{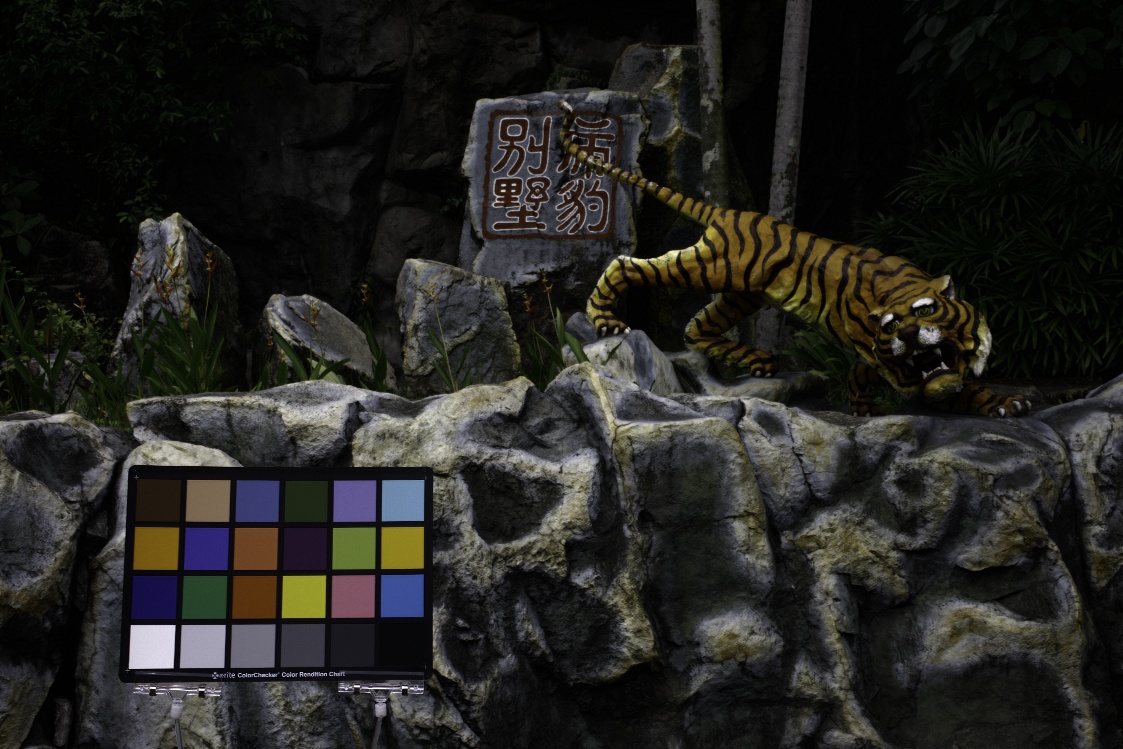} & \includegraphics[width=0.25\linewidth]{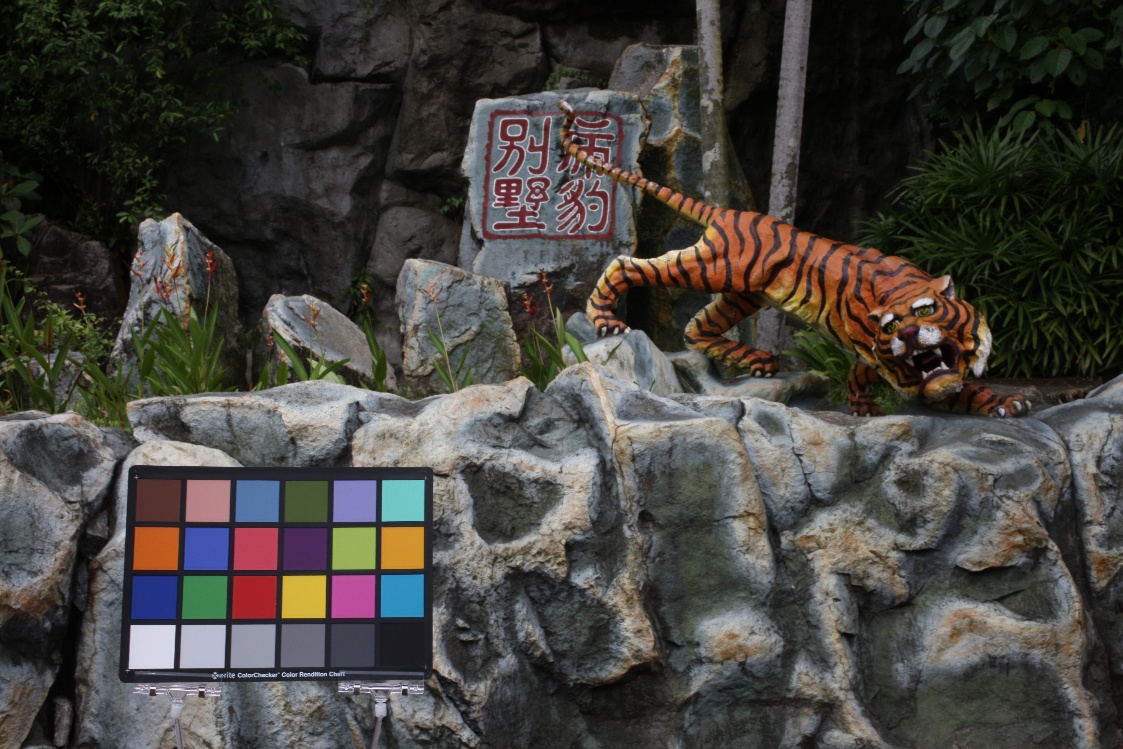} \\
    \includegraphics[width=0.25\linewidth]{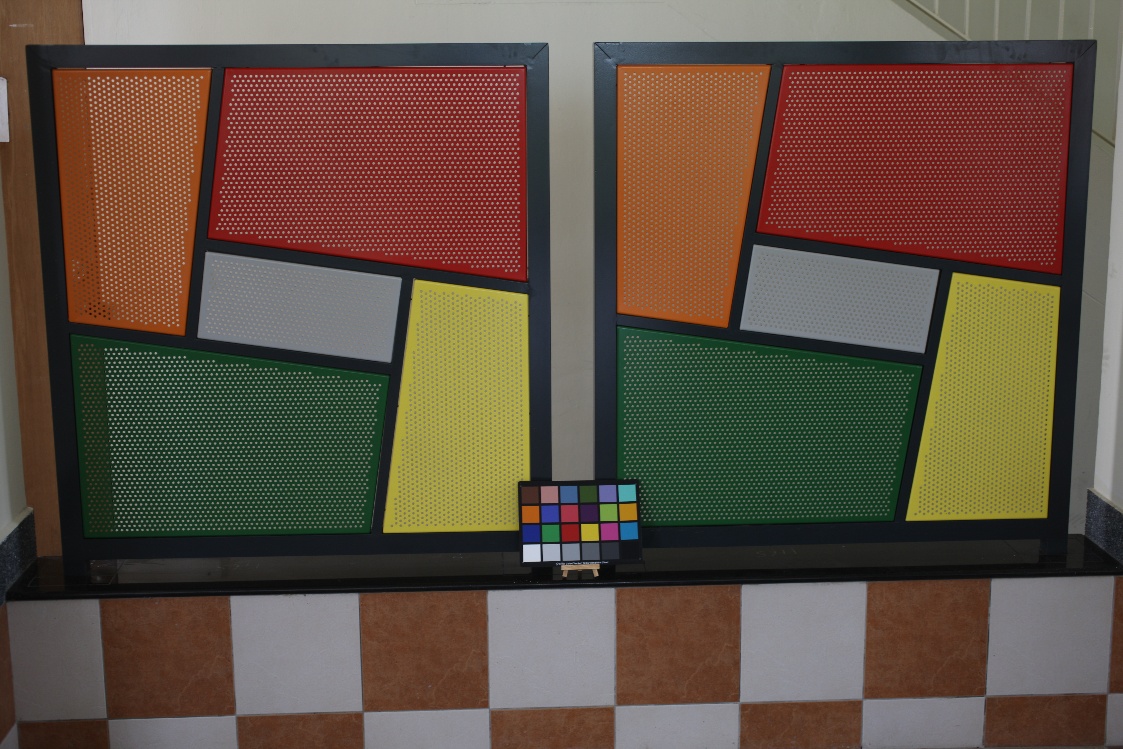} & \includegraphics[width=0.25\linewidth]{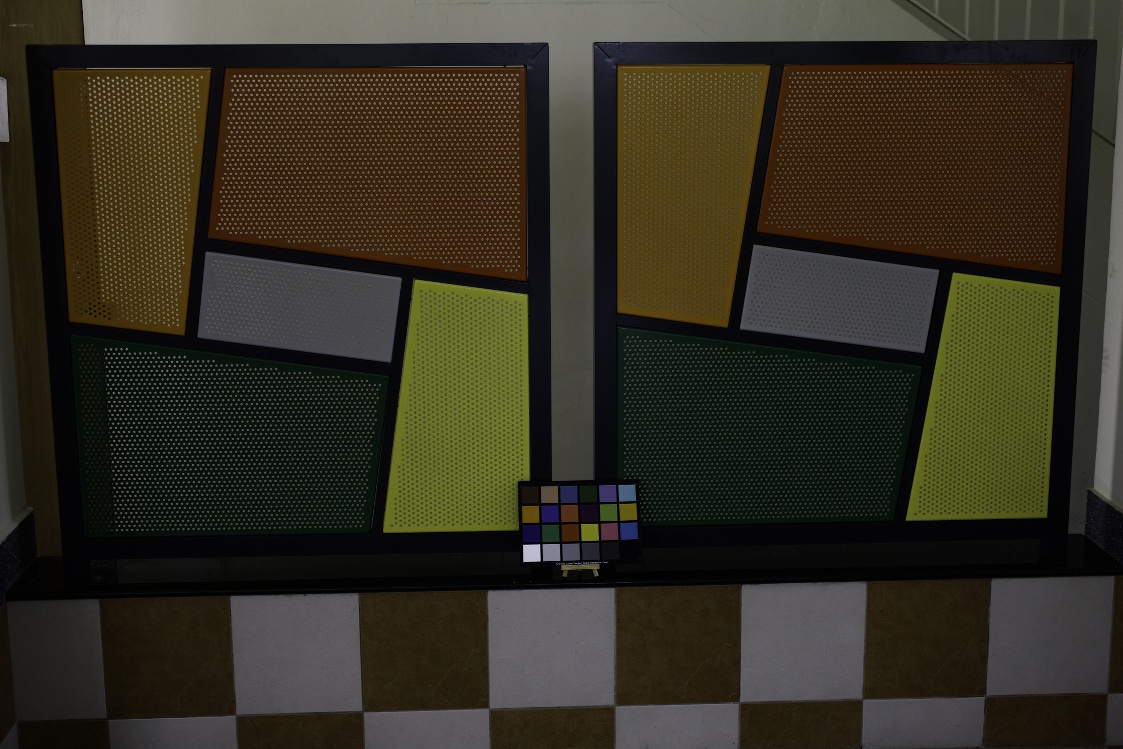} & \includegraphics[width=0.25\linewidth]{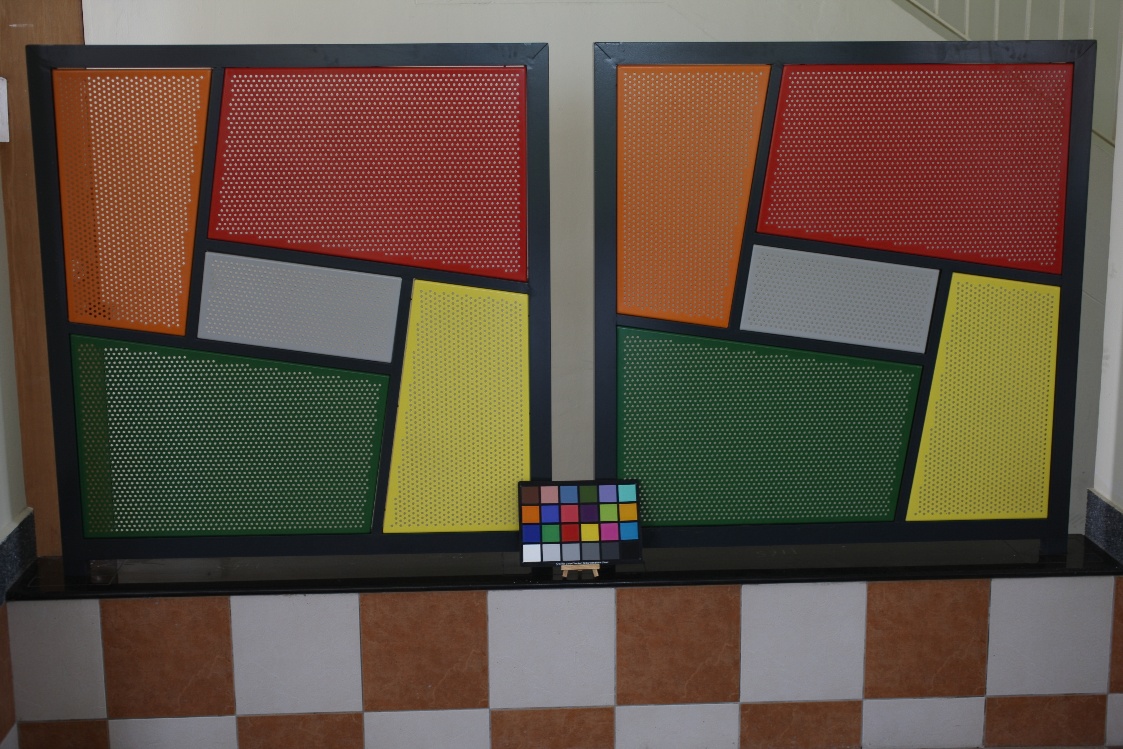} \\
    \includegraphics[width=0.25\linewidth]{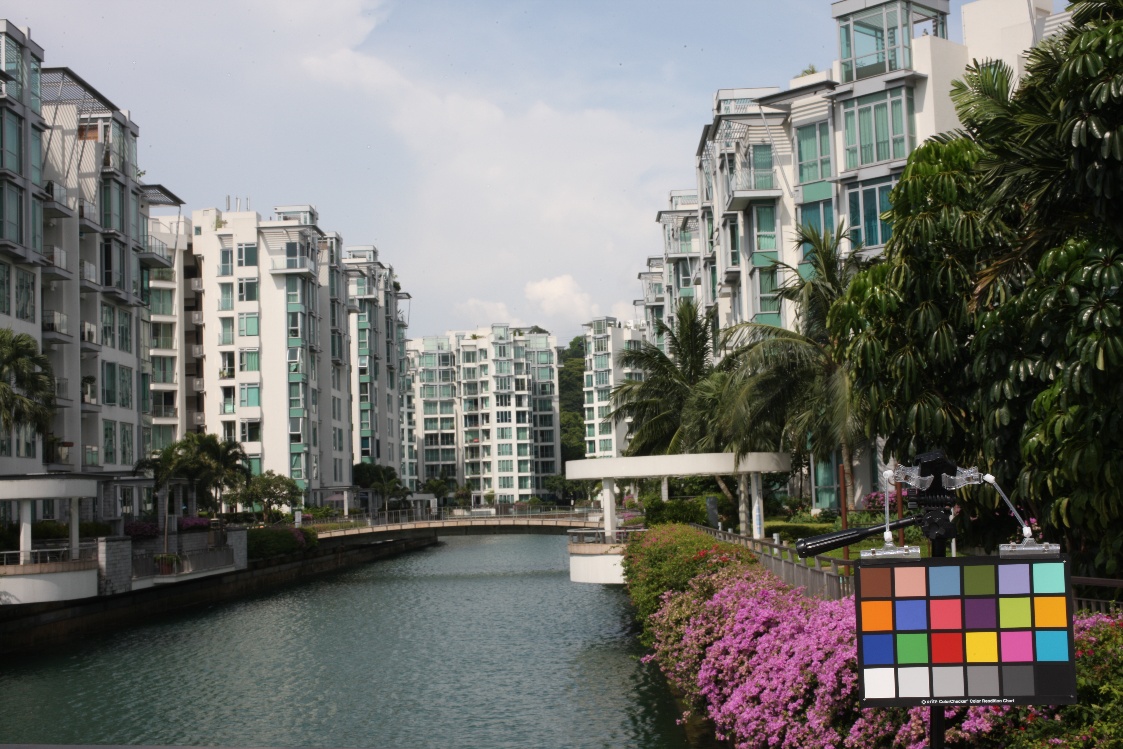} & \includegraphics[width=0.25\linewidth]{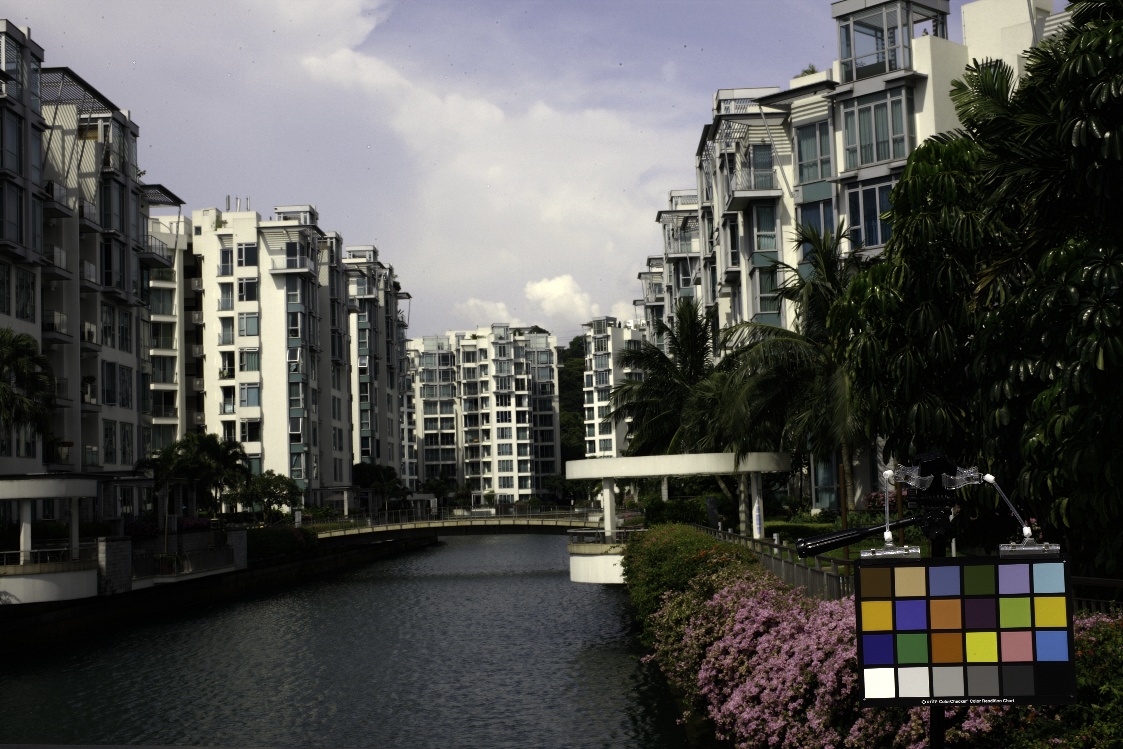} & \includegraphics[width=0.25\linewidth]{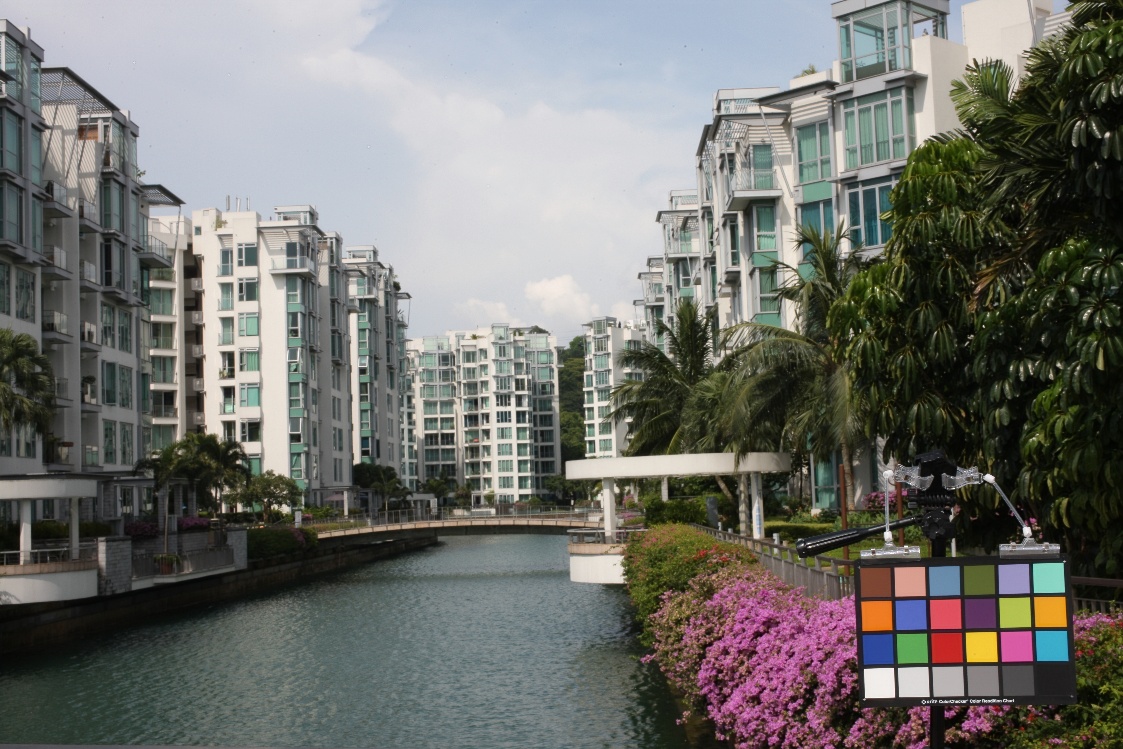} \\
    \includegraphics[width=0.25\linewidth]{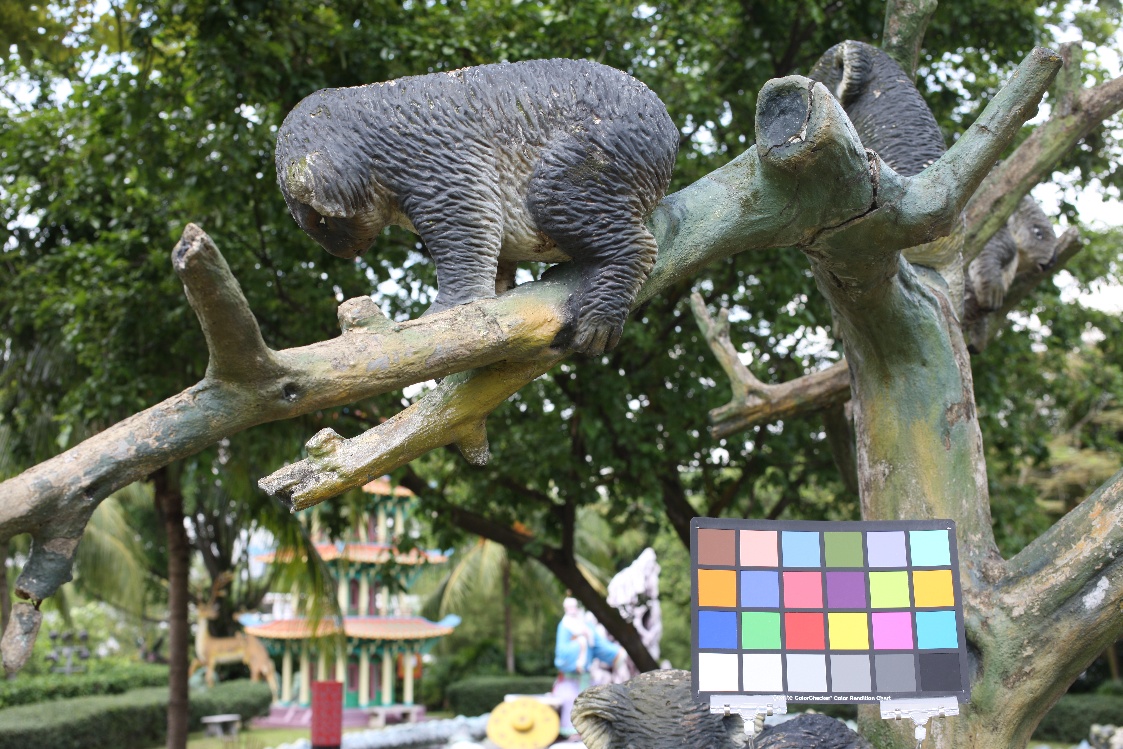} & \includegraphics[width=0.25\linewidth]{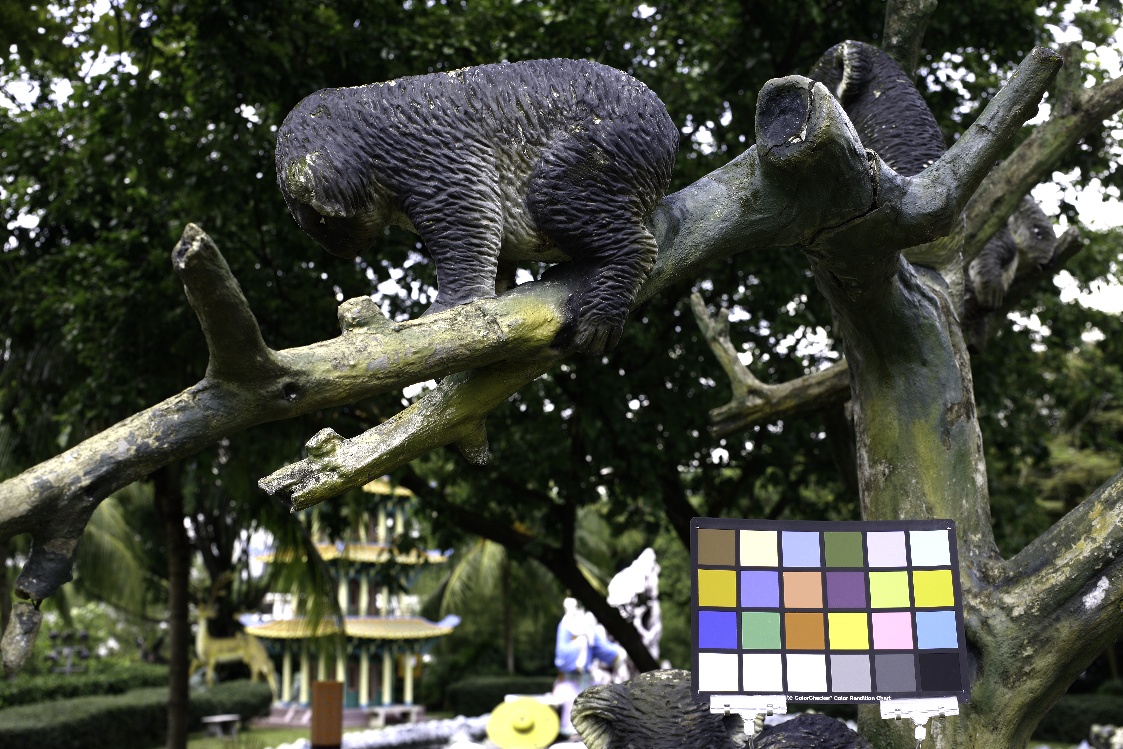} & \includegraphics[width=0.25\linewidth]{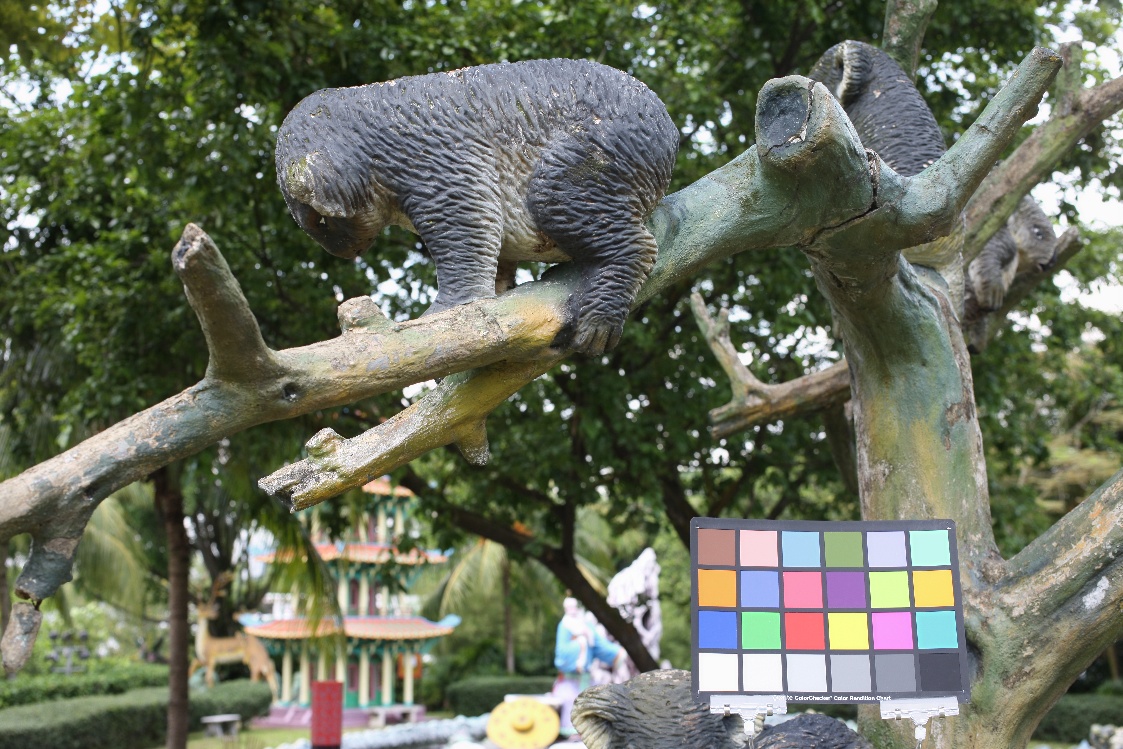} \\
    (A) Input image & (B) Our CIE-XYZ rec. & (C) Our re-rendering
\end{tabular}
}
\caption{Visual comparisons for CIE-XYZ reconstruction and re-rendering. (A) The input sRGB image. (B) CIE-XYZ reconstruction using the proposed method. (C) Our re-rendered output was generated from the reconstructed CIE XYZ image. CIE XYZ images are scaled by a factor of two to aid visualization. The input images are sourced from the NUS dataset \cite{cheng2014illuminant}.}
\label{fig:xyz_states}
\end{figure*}

Using a device-independent linear color space in computer vision applications, such as CIE-XYZ, has proven valuable for multiple tasks \cite{afifi2021cie,brooks2019unprocessing}. The reconstruction of color images in the CIE-XYZ color space from non-linear images is crucial for achieving accurate color representations. However, conventional methods that rely solely on paired CIE-XYZ and sRGB representations face challenges due to the limited availability of large-scale paired image datasets. Acquiring such datasets is often time-consuming, expensive, and not easily scalable, hindering the development of robust color image reconstruction techniques. 
To address the challenges of acquiring labeled datasets at scale (like CIE-XYZ and sRGB pairs), SSL has recently gained significant attention as an alternative paradigm \cite{gui2023survey,Svirsky,eisenberg2024self,rozner2023domain}.
Unlike traditional supervised learning, which relies on externally provided labels, SSL leverages the intrinsic properties of the input data to generate surrogate labels. By exploiting the inherent information in the data, SSL offers a promising avenue to overcome the reliance on paired data and enhance the performance of color image reconstruction.

Most SSL approaches in computer vision focus on image segmentation and classification. Using such SSL techniques has substantially improved performance outcomes in various domains, including medical image analysis and object recognition. For example, SimCLR \cite{chen2020simple} proposed a simple yet effective framework for SSL of visual representations. The SimCLR method used a contrastive learning approach to learn representations that capture the similarity between different views of the same image. SimCLR demonstrated state-of-the-art performance on several benchmark image classification datasets, including CIFAR-10, CIFAR-100 \cite{krizhevsky2009learning}, and ImageNet \cite{russakovsky2015imagenet}.
Utilizing a pre-trained model followed by fine-tuning offers the advantage of requiring less data than training from scratch. However, in certain applications, a suitable pre-trained model may not be available. In such cases, an alternative approach that is both practical and yields comparable results while avoiding the need for large amounts of annotated data required by pre-trained models becomes desirable. This is where SSL comes into play. SSL operates without needing external labels, instead leveraging inherent information in the input data. In this study, we aim to adapt the SSL method to reconstruct CIE-XYZ from non-linear RGB images.

This paper proposes an SSL-based method for reconstructing CIE-XYZ images from non-linear RGB inputs. Our framework aims to mitigate the challenges associated with limited paired data availability. We draw inspiration from successful SSL techniques developed for conventional computer vision tasks. These techniques have been applied in image segmentation and classification tasks, as demonstrated in foundational works such as \cite{liu2021self}. The SimCLR \cite{chen2020simple} framework, in particular, demonstrated remarkable performance in learning visual representations by capturing the similarity between different views of the same image." 

Our methodology employs SSL and includes a pre-task that focuses on the color boards present in the images. We were inspired by the work of \cite{yan2023domain}, who introduced a domain knowledge-guided SSL technique for change detection in remote sensing images, and we adopted a similar approach. The authors in \cite{yan2023domain} utilized prior knowledge of remote sensing indices to direct the learning process and improve change detection capabilities. Similarly, we use prior knowledge of color boards within the images to guide the learning process and enhance the quality of reconstruction.

By leveraging the predetermined colors of patches on the color boards, we develop a self-supervised training paradigm that enables the reconstruction of color patches in the CIE-XYZ color space. This concept can be seen in Fig. \ref{fig:xyz_states}, where we present visual comparisons showcasing CIE-XYZ reconstruction and re-rendering processes.  Incorporating color boards provides an inherent label source for the network during the self-supervised training, eliminating the need for paired CIE-XYZ and sRGB images. To evaluate the effectiveness of our proposed framework, we employ the sRGB2XYZ dataset created by \cite{afifi2021cie}. This dataset, derived from the MIT-Adobe FiveK dataset, offers pairs of sRGB and camera CIE-XYZ images obtained through a camera pipeline. Additionally, we compare our framework's performance against state-of-the-art methods to showcase its color accuracy and reconstruction quality superiority.
\newline

In summary, our contributions are:
\begin{itemize}
    \item We present a new framework that leverages SSL to advance the reconstruction of color images in the CIE-XYZ color space from non-linear RGB inputs. By mitigating the reliance on paired data and drawing inspiration from SSL techniques, our algorithm offers an innovative approach to enhance color image reconstruction.
    \item We benchmark our method and demonstrate its superiority over existing approaches, highlighting its potential impact on various computer vision applications requiring precise color representations.
     \item We further demonstrate that a pre-trained classification network (such as ResNet) can be used to improve performance in CIE-XYZ reconstruction. We, therefore, incorporate such a backbone into our model.
    
\end{itemize}

\section{Related Work}
\label{related work}

Methodologies for de-rendering sRGB images can be classified into two categories: those that incorporate specialized metadata during the capture process and blind methods that do not rely on additional information. Early digital cameras lacked access to the sensor's raw-RGB image, leading to radiometric calibration methods \cite{debevec2008recovering, grossberg2003determining, mitsunaga1999radiometric} focusing on linearizing the sRGB data rather than accurately recovering raw-RGB values. These methods employ simplistic models, such as a primary 1D response function per color channel, to establish a linear relationship between the digital values and scene radiance.
As for the blind methods within the other category, they too can be further divided into two types. The initial type consists of methods attempting to model the parametric relationship that maps from sRGB to some linear state \cite{nguyen2016raw,brooks2019unprocessing}. The second type comprises machine learning methods that aim to learn the color transformation using pairs of sRGB images and their corresponding images used for reconstruction; this includes those in the CIE-XYZ color space \cite{afifi2021cie} and raw-RGB format \cite{nam2022learning}.

In addition to the de-rendering methods that adopt a generic approach, such as those presented in \cite{brooks2019unprocessing, koskinen2019reverse}, which often exhibit limited accuracy due to their inability to model camera-specific operations, there are other works that employ different neural network solutions. For instance, \cite{tang2023bmisp} is a relevant example. These methods assume a standard set of Image Signal Processor (ISP) operations, which hinders their ability to account for individual cameras' unique characteristics.
In contrast, our combination of unsupervised and self-supervised training enables broader generalization, enhancing reconstruction accuracy.
Moreover, while machine learning-based image reconstruction methods can take advantage of the abundance of images available on the internet for training, the absence of the desired image pairs required for supervised training significantly limits their effectiveness and generalization.
Certain studies have incorporated prior knowledge into the learning process in scenarios where the availability of labeled data is limited. Integrating domain-specific information in the training process enhances the reliability of the resulting model \cite{von2021informed}.

In recent years, there has been an increasing interest in incorporating prior knowledge into the training process of machine learning models. As a result, data-driven and knowledge-guided methods have emerged, showing promising improvements in model performance. An innovative approach, proposed by \cite{stewart2017label}, involves supervising neural networks by defining constraints in the output space rather than relying on explicit input-output pairs as training data. These constraints are derived from prior domain knowledge, such as known laws of physics. The authors demonstrate the efficacy of this method on various real-world and simulated computer vision tasks, highlighting its potential to improve the performance of neural networks in practical applications.

Compared to existing approaches in color image reconstruction, our framework offers a distinctive aspect by utilizing guided SSL with color boards, providing an innovative and practical way to enhance CIE-XYZ image reconstruction without the need for extensive paired data. Through evaluations on the sRGB2XYZ dataset \cite{afifi2021cie}, our proposed framework outperforms existing methods, showcasing its potential impact on various computer vision applications requiring precise color representations. Our work contributes to advancing the field of color image reconstruction and demonstrates the viability of SSL in this context.

\begin{figure}[t]
  \begin{subfigure}[b]{0.48\linewidth} 
    \centering
    \includegraphics[width=0.8\linewidth]{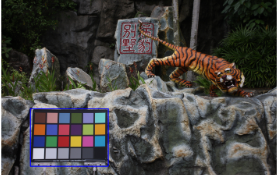}
    \subcaption{}
    \label{subfig:example_roi}
  \end{subfigure}\hfill
  \begin{subfigure}[b]{0.48\linewidth} 
    \centering
    \includegraphics[width=0.8\linewidth]{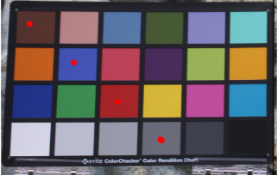}
    \subcaption{}
    \label{subfig:additional_info}
  \end{subfigure}
  \vspace{5pt}
  \caption{The methodology employed for executing self-supervised training using the color boards allows the creation of a pre-task. This pre-task involves retaining and matching the colors of patches located on the color boards with corresponding colors in the CIE-XYZ color space. Sub-figure (\subref{subfig:example_roi}) displays an example image from the NUS dataset \cite{cheng2014illuminant}, where the color board is detected based on the metadata. Additionally, (\subref{subfig:additional_info}) illustrates a visualization of the extra information in the metadata that can be used to sample each color board's color patches during the pre-task.}
  \label{fig:colorboards}
\end{figure}

\section{Proposed Method - Multi-Phase Training Framework for CIE-XYZ Image Reconstruction}

\begin{figure*}[t] 
\centering
\fbox{\includegraphics[width=0.9\textwidth]{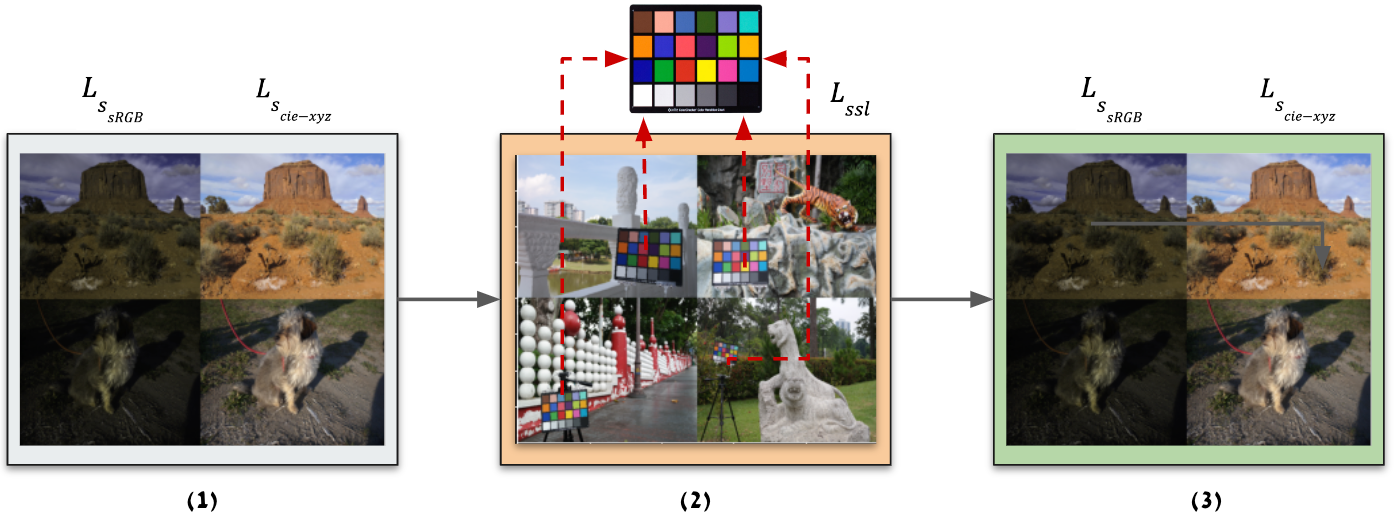}} 
\caption{The training framework consists of three main phases: (1) Supervised training using the sRGB2XYZ dataset \cite{afifi2021cie} to reconstruct CIE-XYZ images with the loss $L_{s_{srgb}}$ for minimizing sRGB image reconstruction error and $L_{s_{cie-xyz}}$ for minimizing CIE-XYZ image reconstruction error, (2) SSL using the NUS dataset \cite{cheng2014illuminant} and color board patches with known CIE-XYZ colors to augment the network, with the loss $L_{ssl}$ for minimizing Delta E 76 discrepancy between reconstructed color board patches and their corresponding ground truth CIE-XYZ values, and (3) An additional phase of supervised training on the sRGB2XYZ dataset for further enhancement and incorporation of supervised data with the same loss functions $L_{s_{srgb}}$ and $L_{s_{cie-xyz}}$.}
\vspace{-0.1 in}
\label{fig:training_steps}
\end{figure*}

The proposed approach follows a multi-phase training scheme for CIE-XYZ image reconstruction, comprising three key training phases with weight adaptation and transfer (see Fig.\ref{fig:training_steps}). First, we use supervised training with pairs of sRGB images and their corresponding CIE-XYZ images to minimize discrepancies between the reconstructed and original images. Then, we perform a self-supervised step based on datasets with sRGB images containing color boards and predetermined colors for color patch regions, enriching the network's learning process. Finally, we perform additional supervised training on a dataset containing sRGB and CIE-XYZ image pairs, refining learned representations and incorporating supervised data for improved performance.

The supervised training phase requires a dataset comprising sRGB images paired with corresponding linear images in the CIE-XYZ color space. To achieve this, we use the sRGB2XYZ dataset \cite{afifi2021cie} derived from the MIT-Adobe FiveK dataset \cite{bychkovsky2011learning}. Creating this dataset involves taking raw-RGB images from the MIT-Adobe FiveK dataset and processing them twice, resulting in both sRGB and CIE-XYZ versions of each image. The authors used the camera pipeline outlined in \cite{abdelhamed2018high} to convert raw-RGB images into the CIE-XYZ color space. This pipeline allowed them to access the CIE-XYZ values by processing the sensor raw-RGB images. The process includes using the color space transformation (CST) matrices provided with the raw-RGB images. The dataset includes ~1200 pairs of sRGB and camera CIE-XYZ images. 
The second type of dataset necessary for the self-supervised part must include sRGB images that contain a color board within the image. Here, we used the dataset presented by \cite{cheng2014illuminant}. The dataset comprises images from 9 commercial cameras, where over 200 images were captured for each camera. The images were taken in natural settings, both indoor and outdoor, with a color board presented within the image. A text file containing the coordinates of the color boards and their corresponding color patches is provided for each image.
Combining these two types of datasets allows the network to learn from both supervised and self-supervised data, contributing to a more robust and comprehensive training reconstruction process.

\subsection{Phase I - Training With Paired Images}
In the first phase, the network undergoes supervised training using the sRGB2XYZ dataset \cite{afifi2021cie}. This dataset contains pairs of sRGB images and their corresponding linear images in the CIE-XYZ color space. The goal is to reconstruct CIE-XYZ images that accurately represent the color information in the input sRGB images. During training, the network's weights are adjusted to minimize the error between the reconstructed CIE-XYZ images and their corresponding ground truth CIE-XYZ images from the dataset.
The supervised training phase is crucial for establishing an anchor for the main task of CIE-XYZ image reconstruction. By learning from the paired sRGB and CIE-XYZ images, the network learns the color relationships and representations required to transform sRGB data into the CIE-XYZ color space accurately.

The loss function (Eq. \ref{ssloss}) used in the supervised training phase:
\begin{equation}
\label{ssloss}
    L_{s} = \lambda|\hat{x}_{xyz} - x^{*}_{xyz}| + |\hat{x}_{srgb} - x^{*}_{srgb}|,
\end{equation}
is derived from \cite{afifi2021cie}, which aims to minimize the mean absolute error (MAE) between the predicted CIE-XYZ image $ \hat{x}_{xyz} $ and its corresponding ground truth $ x^{*}_{xyz}$, as well as the predicted sRGB image $ \hat{x}_{srgb}$ and its corresponding ground truth $ x^{*}_{srgb}$.
The loss enhances color image reconstruction by encouraging the model to capture the meaningful color relationships between the images and produce accurate representations. Additionally, the choice of L1 loss (MAE) over L2 (MSE) is preferred in this context because it can handle outliers and produce more visually pleasing results for color representations.
The value of $ \lambda$ is a weighting factor calibrated by \cite{afifi2021cie}, and we have adopted their value of 1.5 for our implementation.

\subsection{Phase II - Refinement With Color Boards}
In the second phase, the network undergoes SSL using the NUS dataset \cite{cheng2014illuminant}. The dataset contains sRGB images that have color boards positioned within the image. Each color board contains color patches with known colors. The network is now tasked with leveraging this known information to enhance its understanding of the CIE-XYZ color space transformation.
We propose utilizing images containing color boards positioned within them. By doing so, we draw inspiration from the inherent knowledge
 present in the input data, explicitly referring to the known colors of the color patches on the color boards. These predetermined colors act as natural labels that guide the training process, eliminating the need for paired CIE-XYZ and sRGB images and enabling the network to learn the mapping between sRGB and CIE-XYZ color spaces more effectively.
This idea is somewhat analogous to \cite{gidaris2018unsupervised}, who conducted a preliminary task involving the prediction of image rotations. Here, we use the fact that the colors of the patches on the color boards are predetermined and already known (an example can be seen in Fig.\ref{fig:colorboards}).

By utilizing this fact, we established a pre-task wherein the color patches on the color boards were required to correspond to the relevant colors within the CIE-XYZ color space after the CIE-XYZ image reconstruction.

Considering the inherent constraint within the architecture of the proposed neural network, wherein local processing on the image is eliminated, it becomes evident that the network structure requires a transformation, which in this context refers to a matrix multiplication applied to the entire image. Consequently, if a transformation applies to the color board, it applies to all pixels in the image.
The SSL phase is designed to augment the network with additional information using a dataset that may not necessarily comprise image pairs. Instead of relying on external annotations or ground truth labels, the color patches on the color boards act as inherent labels. The network's weights are adjusted to minimize the discrepancy between the actual color values of the color board patches in the CIE-XYZ color space and the color patches reconstructed by the network.
By pre-training on this self-supervised task, the network can learn to associate the color board patches with their corresponding CIE-XYZ colors, thereby gaining valuable knowledge about the color space and improving its ability to reconstruct accurate CIE-XYZ images in the subsequent phases.

 The loss function employed in the self-supervised training phase is based on the Delta E 76 formula \cite{green2023colorimetry}. The primary objective of this loss function is to minimize the Delta E 76 between the reconstructed colors of the color board patches and their corresponding ground truth CIE-XYZ color values. The rationale behind this loss function is to identify a pre-task that does not necessitate pairs of CIE-XYZ and sRGB images, thus allowing for the generalization and enrichment of the training data.
The patch color is determined by sampling the colors inside the patch and representing them as a matrix $C_s$. The matrix $C_s$ has a shape of $n \times 3$, where $n$ is the number of pixels within the mask corresponding to the patch, and each row represents the color value of a pixel in the CIE-XYZ color space. To obtain a single representative color for the patch, the 75th percentile value of $C_s$ is taken as ${\mathcal{F}}_{q_{75}}$ and used to obtain $C_{sq}$, defined as:
\begin{equation}
    C_{sq} = \mathcal{F}_{q_{75}}(C_{sq}),
\end{equation}
which is the reconstructed patch color in the CIE-XYZ color space.
To utilize the Delta E method, colors must be converted to the CIELAB color space \cite{schanda2007colorimetry}. In this study, the reconstructed patch color $C_{sq}$ is transformed into $C_{rpc}$ using the conversion method outlined in \cite{pub2004technical}. The conversion function is denoted as $\mathcal{F}_{xyz->lab}$ and is used to convert from CIE-XYZ to CIELAB, as shown below:
\begin{equation}
C_{rpc} = \mathcal{F}_{xyz->lab}(C_{sq}).
\end{equation}

The Delta E method computes the color difference between the reconstructed patch color and the ground truth color of the $i$-th patch in the LAB color space $C_{gtpc}$, which is provided as a property of the color checker, the color checker is a color board that complies with international standards and is widely used in camera calibration and color correction \cite{gong2016color}. The Delta E for the $i$-th patch is calculated using the following equation:
\begin{multline}
\Delta{E}_{i} = \sqrt{(C_{rpc}^{L}- C_{gtpc}^{L})^2} \\
+ \sqrt{(C_{rpc}^{a} - C_{gtpc}^{a})^2} + \sqrt{(C_{rpc}^{b} - C_{gtpc}^{b})^2}.
\end{multline}

Finally, the self-supervised loss function part is denoted as:
\begin{equation}
    L_{ssl} = \mathcal{F}_m(\{\Delta{E}\}_{i=0}^{i=n}),
    \label{loss_ss}
\end{equation}
where $\mathcal{F}_m$ is the mean operator, and $n$ is the number of color patches in the color board.

During the self-supervised training, the loss function combines two components - the self-supervised loss described in Equation \ref{loss_ss} and the sRGB part of the supervised loss.
\begin{equation}
    L_{sslt} = \delta |\hat{x}_{srgb} - x^{*}_{srgb}| + L_{ssl}.
\end{equation}
Combining the above two terms balances the two aspects of the network's learning process. While the self-supervised loss facilitates the network in acquiring valuable insights into the transformation between the sRGB and CIE-XYZ color spaces, the sRGB part of the supervised loss ensures that the network accurately reconstructs the original sRGB image. 

To balance the self-supervised learning and the supervised reconstruction tasks, we introduced a trainable parameter, denoted as $\delta$, into our loss function. This addition empowers us to dynamically calibrate the interplay between self-supervised and supervised losses, thus enabling the network to progressively refine its focus on these dual objectives during the learning process. The $\delta$ parameter, a singular scalar, effectively controls the relative significance of the two types of loss elements. By incorporating $\delta$ as a trainable parameter, the network can find the best balance between grasping color space insights and achieving accurate sRGB image reconstruction.
\subsection{Phase III - Final Supervised Refinement}
In the final phase, the network undergoes another round of supervised training using the sRGB2XYZ dataset \cite{afifi2021cie}. Similar to Phase I, this phase involves training the network on pairs of sRGB images and their corresponding linear images in the CIE-XYZ color space.
The purpose of this final supervised training phase is twofold. Firstly, it aims to enhance the information learned in the previous self-supervised phase by fine-tuning the representation learned by the network. Secondly, it addresses any errors or biases that might have occurred during the self-supervised training process.

By using a multi-phase training approach, the network can benefit from both supervised and semi-supervised learning. The supervised training provides a solid foundation for the main task of CIE-XYZ image reconstruction, while the SSL with color boards augments the network's understanding of the CIE-XYZ color space. Combining these phases enables the network to produce more accurate and reliable CIE-XYZ image reconstructions.

\section{Experimental results}

Our approach is evaluated on the benchmark proposed by \cite{afifi2021cie}, utilizing the test set of the sRGB2XYZ dataset. The benchmark serves to validate the efficacy of the proposed framework in the mapping of camera-generated sRGB images to CIE-XYZ and the processing of CIE-XYZ images back to sRGB.

\subsection{Implementation Details}
\begin{figure*}[t]
\centering
\fbox{
  \includegraphics[width=0.8\linewidth]{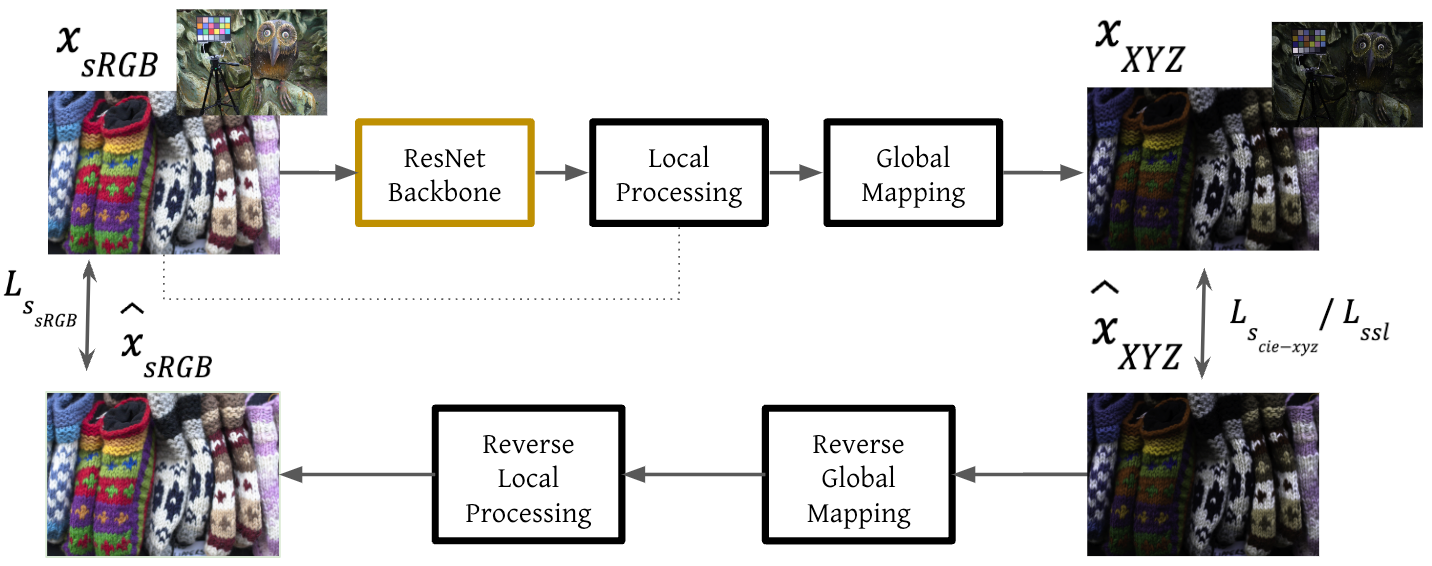}
}
\caption{The CIE-XYZ image pipeline network from \cite{afifi2021cie} with the ResNet pre-trained ImageNet backbone. The neural network architecture used in this study aims to emulate the camera imaging pipeline. It comprises two sub-networks that model both the global and local processing parts. The network's backbone, responsible for feature extraction and representation learning, is initialized using a pre-trained ResNet50 based on the architecture proposed in \cite{he2016deep}. This ResNet50 backbone was initially trained on the ImageNet dataset \cite{russakovsky2015imagenet}, a large-scale dataset with various annotated images. The pre-trained backbone is integrated into the network before the local processing convolutional neural network (CNN), allowing it to benefit from the learned features during pre-training. The pre-trained weights of the backbone are then fine-tuned during subsequent training phases, where the network learns to perform the specific task at hand, resulting in an efficient and effective image-processing pipeline.}
\label{resnet}
\end{figure*}

Our neural network architecture is based on the network described in \cite{afifi2021cie}, which aims to emulate the camera imaging pipeline. The architecture comprises two sub-networks that model global and local processing parts. This architecture is employed across all three training phases, with the weights transferred from one phase to another.
Pre-training is a prevalent approach in computer vision, wherein the backbone of object detection and segmentation models is often initialized using supervised ImageNet pre-training. Our research explored another innovation, which involves using a pre-trained backbone before the local processing CNN, as depicted in Fig.\ref{resnet}. Specifically, we utilized a pre-trained ResNet50 based on the architecture proposed in \cite{he2016deep} that was trained on the ImageNet dataset \cite{russakovsky2015imagenet}. This choice of using ResNet50 as the pre-trained backbone was motivated by its proven effectiveness in a wide range of computer vision tasks, demonstrated by its outstanding performance in various benchmarks. Moreover, ResNet50's depth and skip connections facilitate feature extraction at multiple scales, which is particularly beneficial for tasks like image reconstruction.

The split of training phases into supervised and self-supervised components is based on the availability and characteristics of the datasets and provides several benefits. The first and third components, which are supervised, rely on the sRGB2XYZ dataset \cite{afifi2021cie}, consisting of 971 pairs of sRGB images and their corresponding linear CIE-XYZ images. This dataset was suitable for the supervised training as it provided ground truth pairs, enabling the network to learn the color transformation accurately. On the other hand, the self-supervised part of the training utilized a dataset introduced by \cite{cheng2014illuminant} containing sRGB images with color boards. We integrated this dataset into our training process, enabling the network to leverage inherent labels from the color board patches and learn more about the CIE-XYZ color space transformation without relying solely on paired data. The split was chosen to ensure that both supervised and self-supervised learning components were optimally trained with appropriate datasets for their respective tasks. 

During the supervised training, the network was trained with randomly extracted patches of size $256 \times 256$ from the training set, with a mini-batch size of 4. Additionally, scaling and reflection augmentations were employed on the extracted patches to enhance the training process.
The stages of the framework were trained for 300 epochs each, employing the Adam optimizer \cite{kingma2014adam} with $\beta_1$ = 0.9 as the gradient decay factor and $\beta_2$ = 0.999 as the squared gradient decay factor. A learning rate of $10^{-4}$ was utilized, with a decay factor of 0.5 every 75 epochs to enable convergence to a lower minimum. Additionally, to prevent overfitting, we incorporated an $L2$ regularization into our loss function in Eq.{\ref{ssloss} with a regularization weight of $\lambda_{reg}$ = $10^{-3}$.
The choice of the parameters aligns with the parameters used in \cite{afifi2021cie}, which have been demonstrated to be effective for similar tasks.

The proposed method imposes an inherent constraint on the neural network's architecture, specifically requiring global processing across the entire image instead of local operations. This constraint ensures that the network can perform transformations encompassing the whole image, allowing it to be compatible with the proposed framework for CIE-XYZ image reconstruction. Moreover, this global processing capability is essential for successfully utilizing the color boards-based pre-train task.

\begin{table*}[h!t]
\begin{center}
  \scriptsize
  \begin{tabular}{|r|l|l|l|l|l|l|l|l|l|l|l|l|l|}
  \hline
  Method  & \multicolumn{4}{|c|}{sRGB $\rightarrow$ XYZ}  & \multicolumn{4}{|c|}{Rec. XYZ $\rightarrow$ sRGB}   & \multicolumn{4}{|c|}{GT XYZ $\rightarrow$ sRGB}    \\\hline
    & Avg. & Q1 & Q2 & Q3 & Avg. & Q1 & Q2 & Q3 & Avg. & Q1 & Q2 & Q3 \\ \hline
    Standard \cite{anderson1996proposal,ebner2007color}      & 21.84        & 16.88 & 20.91 & 25.24  & - & - & - & - & 22.22 & 19.19 & 21.79 & 24.37 \\
    Unprocessing \cite{brooks2019unprocessing} & 22.19 & 19.31 & 22.12 & 24.75 & 37.72 & 37.78 & 40.56 & 41.88 & 18.04 & 15.67 & 17.79 & 20.02 \\
    Afifi et al. \cite{afifi2021cie} & 29.66  & 23.77 & 29.57 & 34.71 & 43.82 & 41.43 & 43.94 & 46.58 & 27.44 & 23.57 & 28.32 & 30.88 \\
    SEL-CIE & 30.38& 24.51& 30.46& 35.16& 46.43  & 42.49 & 46.04 & 50.54 & 27.87& 23.86& 28.8& 31.49 \\
    SEL-CIE-RB & 32.11 & 27.49 & 32.02 & 36.49 & 44.51 & 41.64 & 44.72 & 47.79 & 27.94 & 24.11 & 29.04 & 31.55           \\\hline
  \end{tabular}
  \end{center}
  \caption{PSNR comparison across various methods: first, sRGB to CIE-XYZ reconstruction using ground truth CIE-XYZ; second, sRGB image reconstruction from reconstructed CIE-XYZ with ground truth being the original sRGB image; and finally, PSNR comparison between the network's reconstruction (input: ground truth CIE-XYZ, output: reconstructed sRGB) and the corresponding ground truth sRGB image. The proposed SEL-CIE method surpasses existing methods in all metrics. Additionally, integrating a pre-trained ResNet50 backbone further improves performance (SEL-CIE-RB).}
  \label{tab:conversion-comparison}
\end{table*}

\subsection{Evaluation Metrics}
Our framework's capability to "unprocess" sRGB images to CIE-XYZ and reconstruct them from CIE-XYZ back to sRGB is verified and demonstrated. To evaluate our framework's mapping to sRGB, we conduct experiments using our reconstructed CIE-XYZ results and ground-truth CIE-XYZ images as the starting points. For evaluation purposes, we compare our approach with the supervised training method in \cite{afifi2021cie} and the standard CIE-XYZ mapping in \cite{anderson1996proposal} and \cite{ebner2007color}, which uses a simple 2.2 gamma tone curve. Additionally, we compare our results with the unprocessing technique (UPI) in \cite{brooks2019unprocessing}, which provides a proxy for the major procedures of the camera pipeline. We compare our results with UPI obtained at the CIE-XYZ stage to ensure a fair evaluation.

Following the proposed benchmark in \cite{afifi2021cie}, Table \ref{tab:conversion-comparison} shows peak-signal-to-noise ratio (PSNR) results averaged over the 244 unseen testing images from the sRGB2XYZ dataset. The terms Q1, Q2, and Q3 represent the first (lowest), second (median), and third quartile, correspondingly, of the PSNR (Peak Signal-to-Noise Ratio) values achieved by each approach.

Table \ref{tab:conversion-comparison} illustrates that our proposed method (SEL-CIE) has yielded superior results across all evaluated metrics. Furthermore, incorporating a pre-trained ResNet50 backbone (SEL-CIE-RB) has further improved the performance in the case of sRGB to CIE-XYZ transformation.

In our comprehensive evaluation, we employed the Structural Similarity Index (SSIM) \cite{wang2004image} as a robust metric to gauge the degree of similarity between the reconstructed CIE-XYZ images and their corresponding ground truth images extracted from the sRGB2XYZ dataset \cite{afifi2021cie}. SSIM takes into account various image attributes such as luminance, contrast, and structural features, providing a holistic assessment of image similarity. A higher SSIM index signifies a more significant resemblance between the images.

Table \ref{tab:ssim-results}, situated below, provides a summarized view of the average SSIM values computed across the testing images for each of the different models used in our evaluation. It serves as a valuable reference point for understanding the performance of these models in preserving image fidelity.

\begin{table}[htbp]
    \centering
    \begin{tabular}{|c|c|}
    \hline
    Method & Average SSIM \\ \hline
    SEL-CIE-RB & 0.9408 \\ \hline
    SEL-CIE & 0.9363 \\ \hline
    Afifi et al. \cite{afifi2021cie} & 0.9338 \\ \hline
    \end{tabular}
\caption{Comparison of Average Structural Similarity Index (SSIM) Results for Image Reconstruction Methods}
\label{tab:ssim-results}
\end{table}

These results highlight the superiority of the SEL-CIE-RB model in preserving both structural and perceptual similarity compared to previous models.

%
%

\section{Conclusion and Future Work}
This paper has presented a framework that harnesses the power of self-supervised learning (SSL) to enhance the reconstruction capabilities of CIE-XYZ images from their corresponding non-linear sRGB images.
By reducing our dependency on paired data and leveraging insights derived from SSL techniques, our framework showcased its ability to reduce errors in CIE-XYZ image reconstruction and sRGB image re-rendering. This accomplishment holds potential for various computer vision applications that demand the use of linear color representations.
Furthermore, our approach has excelled compared to existing methods, emphasizing its capability to advance the field of image processing. This suggests that our framework could be valuable in diverse domains, from medical imaging to color-sensitive computer vision tasks.
In addition to the SSL techniques, we have integrated a pre-trained ResNet50 backbone into our framework, resulting in an even more refined transformation process from sRGB to CIE-XYZ. This integration underscores the versatility of our approach and its potential to enhance color image reconstruction further. 

Future research can explore ways to enhance our framework's generalization capabilities. While it has shown promise, there may be scenarios and datasets where its performance could be further improved. Investigating techniques to adapt the SSL model to varying imaging conditions and scene complexities is important. Additionally, collecting a dataset with color boards from a diverse range of camera types, including smartphone cameras, can contribute significantly to the model's generalization. This broader dataset can help the model adapt to different camera manipulations and sensor characteristics, further improving its robustness. As an example of exploring ways to enhance generalization capabilities, one potential scenario for future research is collaborating with companies in the medical field that develop camera-based products aimed at achieving color normalization and standardization. This approach could be an exciting exploration avenue to train our framework with their data, which may include images with color boards, enabling the algorithm to work independently on different cameras, regardless of the specific processing performed in each camera. Moreover, there exists an opportunity to explore the practical implementation of our suggested framework within the field of medical imaging, particularly in contexts necessitating color standardization and normalization across different cameras. Medical scenarios often demand exceptional precision, particularly when dealing with color accuracy. Tasks such as analyzing wound tissues or interpreting color variations in diagnostic tests critically hinge upon obtaining precise color information. Adapting the usage of our framework to address these specific requirements in medical imaging could yield substantial benefits. This research direction can pave the way for improved diagnostic accuracy and heightened reliability in medical assessments by addressing the unique challenges related to standardization and normalization within the medical domain.

\nolinenumbers

\end{document}